\documentclass[aps, prd, amsmath, preprintnumbers, floats, floatfix, twocolumn,superscriptaddress, nofootinbib, showpacs,eqsecnum]{revtex4-1}
\usepackage{graphicx}
\usepackage{amssymb}
\usepackage{color}
\usepackage{enumerate}

\newcommand{\be}{\begin{eqnarray}}
\newcommand{\ee}{\end{eqnarray}}
\newcommand{\rar}{\rightarrow}

\begin{document}

\title{Can an astrophysical black hole have a topologically non-trivial event horizon?}
\author{Cosimo Bambi}
\email{cosimo.bambi@ipmu.jp}
\affiliation{Institute for the Physics and Mathematics of the Universe, 
The University of Tokyo, Kashiwa, Chiba 277-8583, Japan}
\affiliation{
Arnold Sommerfeld Center for Theoretical Physics,
Ludwig-Maximilians-Universit\"at M\"unchen, 80333 Munich, Germany}

\author{Leonardo Modesto}
\email{lmodesto@perimeterinstitute.ca}
\affiliation{Perimeter Institute for Theoretical Physics, 
Waterloo, Ontario N2L 2Y5, Canada}

\date{\today}

\preprint{IPMU11-0058}

\begin{abstract}
In 4-dimensional General Relativity, there are several theorems 
restricting the topology of the event horizon of a black hole. 
In the stationary case, black holes must have a spherical horizon, 
while a toroidal spatial topology is allowed only for a short time.
In this paper, we consider spinning black holes 
inspired by Loop Quantum Gravity and by alternative theories of
gravity. We show that the spatial topology of the event horizon of 
these objects changes when the spin parameter exceeds a critical 
value and we argue that the phenomenon may be quite common
for non-Kerr black holes. Such a possibility may be relevant in 
astrophysics, as in some models the accretion process 
can induce the topology transition of the horizon. 
\end{abstract}


\maketitle

%%%%%%%%%%%%%%%%%%%%%%%%%%%%%%%

\section{Introduction}

The study of black hole (BH) uniqueness theorems started more 
than forty years ago and it is still a very active research 
field~\cite{review}. In 4-dimensional General Relativity, the 
Hawking's theorem ensures that the spatial topology of the event 
horizon must be a 2-sphere in the stationary case, under the main 
assumptions of asymptotically flat space-time and validity of the 
dominant energy condition~\cite{hawking}. Technically, the event 
horizon of a BH is defined as the boundary of the causal past of 
future null infinity. The spatial topology of the event horizon at a 
given time is the intersection of the Cauchy hypersurface at that 
time with the event horizon. According to the topological censorship 
theorem, in a globally hyperbolic and asymptotically flat space-time, 
any two causal curves extending from past to future infinity are 
homotopic~\cite{tct}. As a BH with a toroidal spatial topology would 
violate this theorem, the hole must quickly close up, before a light 
ray can pass through~\cite{tedj}. Interestingly, numerical simulations 
find that toroidal horizons can form, but they exist for a short time, 
consistently with the topological censorship theorem~\cite{teuk}.

In the vacuum, the only stationary and axisymmetric BH solution
of the Einstein's equations in a 4-dimensional and asymptotically
flat space-time is given by the Kerr geometry, which is completely
specified by two parameters: the mass $M$ and the spin angular
momentum $J$ -- instead of $J$, it is more commonly used the 
spin parameter $a = J/M$ or the dimensionless spin parameters
$a_* = J/M^2$. While the Kerr metric may be a solution even in 
other theories of gravity, in general there is not a similar uniqueness 
theorem~\cite{kerr}. Unfortunately, for the time being our knowledge 
of non-Kerr BH solutions in alternative theories of gravity is definitively 
limited. In most cases, analytic or numerical metrics are 
known only for non-rotating BHs. Approximated solutions in 
the slow-rotation limit have been obtained in Einstein-Gauss-Bonnet-dilaton
(EGBd) gravity~\cite{slow1} and in Chern-Simons modified 
gravity~\cite{slow2}. The only 4-dimensional non-Kerr spinning
BH example of a specific gravity theory is given by a numerical 
metric in EGBd gravity, recently found in~\cite{egbg}.
On the other hand, from the observational point of view, fast-rotating 
BHs are the most interesting, as deviations from the Kerr metric 
are typically more evident and we would have more chances to
test the model with astrophysical data~\cite{rev-b}.

In this paper, we discuss two examples of spinning BHs in theories
beyond General Relativity, proposed respectively in~\cite{bh2} and \cite{dim}.
In both cases, the two metrics have not been obtained by solving
specific field equations, but the fact they have an analytic form for
arbitrary values of the spin parameter is very useful. The BH proposed
in~\cite{bh2} is inspired by Loop Quantum Gravity and the deformations 
from the Kerr geometry are encoded in a polymeric function $P$ and 
in a Plank scale parameter $a_0$. The BH proposed in~\cite{dim}
is instead a phenomenological metric and may be seen as a BH solution 
in an unknown alternative theory of gravity\footnote{These metrics can 
be solutions of particular non-local 
generalizations of the Einstein's equations,
as suggested in~\cite{ModestoMoffatNico,SRQG}.
The idea is to replace the Einstein's equations with the following 
set of equations of motion
$$  R_{\mu \nu} -\frac{1}{2} g_{\mu \nu} R = 
8 \pi G_N \mathcal{O}( \Box/\Lambda^2) T_{\mu \nu} \, ,$$ 
where $\mathcal{O}(\Box/\Lambda^2)$ is a generic non-local function of the 
covariant D'Alembertian operator and $\Lambda$ is the energy scale
of the modified gravity~\cite{SRQG}.  }.
Both BHs have been obtained through a Newman-Janis 
transformation of a static solution.
The key point of the two metrics is that there are no restrictions on the 
values of the spin parameter $a_*$. Interestingly, for high values of 
$a_*$ they present similar features. If the BH is more prolate than 
the predictions of General Relativity, when it rotates fast
the spatial topology of the event horizon changes from a
2-sphere to two disconnected 2-spheres. If the BH is more oblate,
one finds a toroidal horizon or something very similar to a toroidal
horizon. Our guess is that fast-rotating BHs with non-trivial topology
are not peculiar predictions of these two solutions, but that they may
be relatively common in the case of deviations from the Kerr geometry.

\section{Black holes inspired by Loop Quantum Gravity}

Loop Quantum Gravity is a generally covariant and non-perturbative
quantization of General Relativity~\cite{rov}. BHs in Loop
Quantum Gravity have been studied only very recently and it has 
been shown that the central singularity can be solved~\cite{bh1,benedictis}. 
Rotating BHs have been obtained in~\cite{bh2} through a
Newman-Janis transformation. In Boyer-Lindquist coordinates, the 
loop-modified Kerr metric reads~\cite{bh2}
\be\label{eq-metric}
g_{tt} &=& - \frac{(\rho^2 + 2 M P r)^2 \Delta}{\rho^4 \Sigma}
 \, , \nonumber\\
g_{t\phi} &=& - \frac{a \sin^2\theta (\rho^2 + 2 M P r)^2 
(\Sigma - \Delta)}{\rho^4 \Sigma} \, , \nonumber\\
g_{\phi\phi} &=& \sin^2\theta \left[ \Sigma
+ \frac{a^2 \sin^2\theta (\rho^2 + 2 M P r)^2 (2 \Sigma 
- \Delta)}{\rho^4 \Sigma} \right] \, , \nonumber\\
g_{rr} &=& \frac{ (\rho^2 + 2 M P r)^2 \Sigma}{\rho^4 \Delta 
+ a^2 \sin^2\theta(\rho^2 + 2 M P r)^2} \, , \nonumber\\
g_{\theta\theta} &=& \Sigma \, ,
\ee
where
\be
\rho^2 &=& r^2 + a^2 \cos^2\theta \, , \nonumber\\
\Delta &=& r^2 - 2 M (1 + P^2) r + 4 M^2 P^2 + a^2 \cos^2\theta \, ,
\ee
$P$ is the polymeric function
\be
P = \frac{\sqrt{1 + \gamma_I^2 \delta^2} - 1}{\sqrt{1 
+ \gamma_I^2 \delta^2} + 1} \, ,
\ee
and $\gamma_I$ and $\delta$ are respectively the Immirzi parameter
and the ``polymeric parameter''. In principle, $P$ can be 
either positive or negative, because $\gamma_I$ may be complex.
As for $\Sigma$, there is more than one possibility, because of
some ambiguities in the procedure to get the metric. There are
indeed at least two natural complexifications in the Newman-Janis
construction procedure~\cite{bh2}:
\begin{eqnarray}
\label{type1}
{\rm Type \,\, I} \, : \,\,\,\,\,\,  
\Sigma & = & \rho^2 + B^2/r^2 \, , \\
{\rm Type \,\, II} \, : 
\,\,\,\,\,\,    \Sigma & = & \rho^2  + B^2/\rho^2 \, ,
\label{type2}
\end{eqnarray}
where $B$ is the ``bounce constant'' and has been fixed in two 
different ways in previous papers for the spherically symmetric 
solution~\cite{bh1,Mann}:
\begin{eqnarray}
\label{typea}
{\rm Case \,\, a} \, : \,\,\,\,\,\,
B & = & a_0 \, , \\
{\rm Case \,\, b} \, : 
\,\,\,\,\,\,
B & = & (2 M P)^2 \, .
\label{typeb}
\end{eqnarray}
Here $a_0 \sim L_{Pl}^2$ is the minimum area ($L_{Pl} 
\sim 10^{-33}$~cm is the Planck length). Since in this paper we 
are interested in astrophysical BHs, we can neglect 
Planck-scale structures and we assume $a_0 = 0$. We have thus
three slightly different solutions: Ia-IIa, Ib, and IIb.
However, all our conclusions are independent of the
ambiguities related to the $\Sigma$.

\begin{figure*}
\par
\begin{center}
\includegraphics[type=pdf,ext=.pdf,read=.pdf,width=6cm]{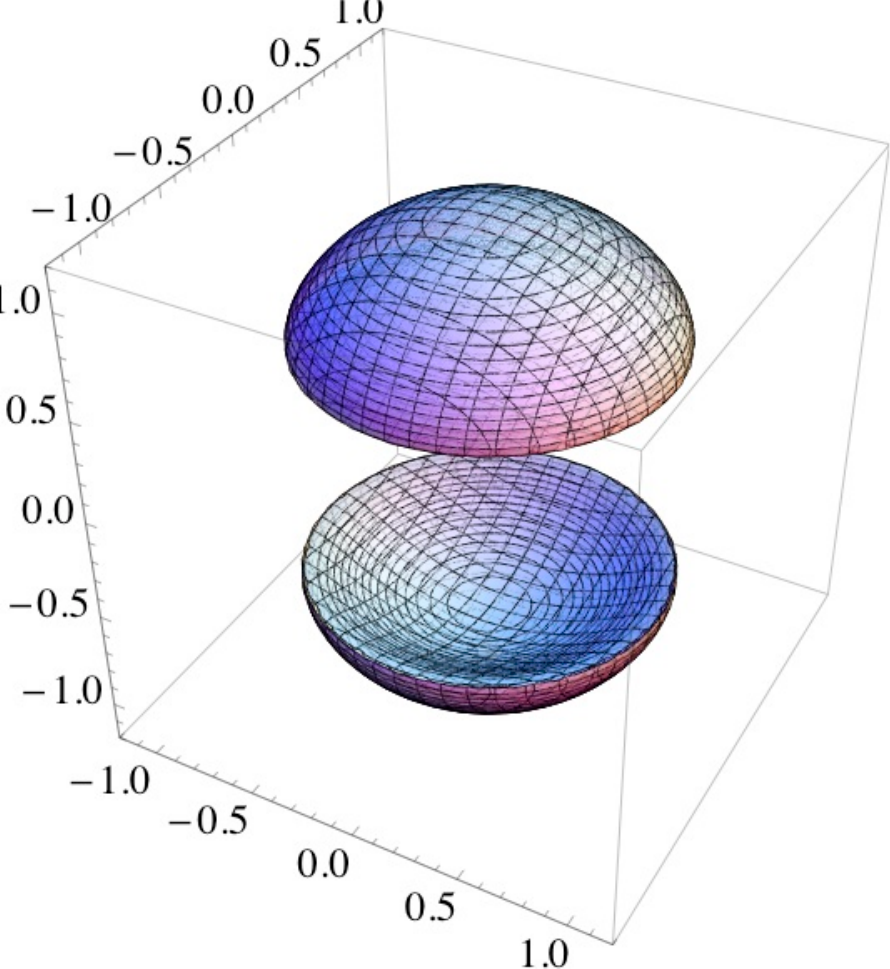}
\hspace{1cm}
\includegraphics[type=pdf,ext=.pdf,read=.pdf,width=6cm]{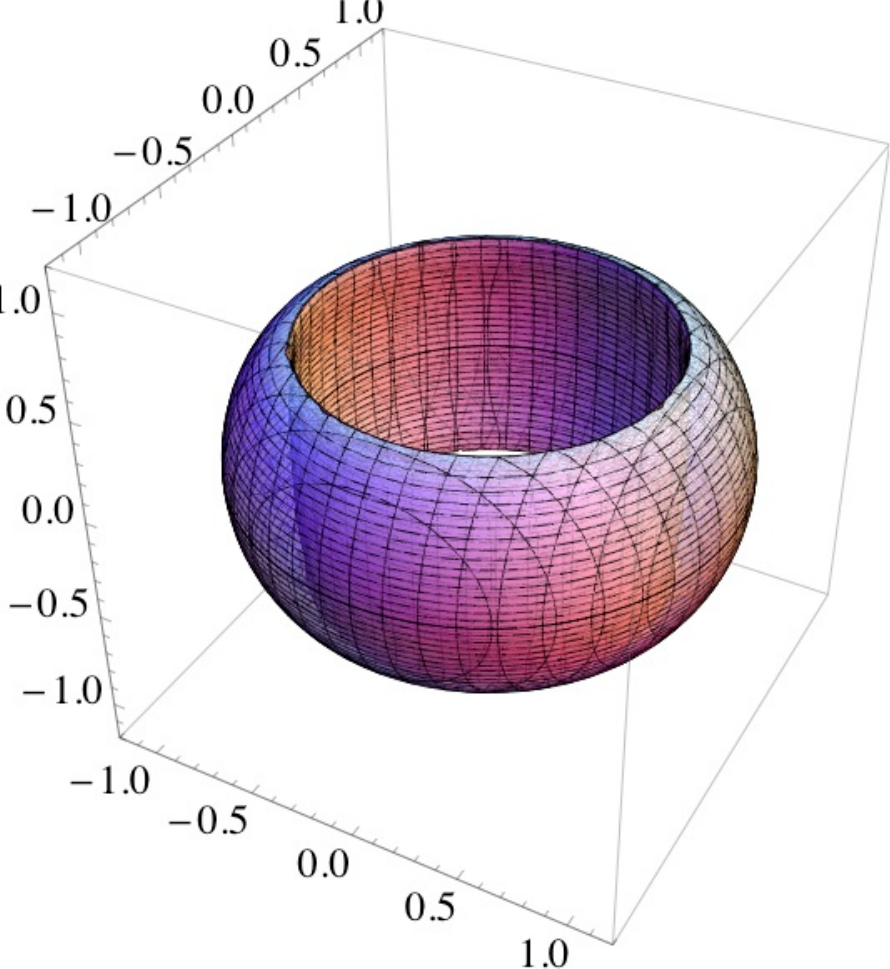}
\par
\vspace{0.3cm} 
\caption{Event horizons of loop-inspired black holes.
Left panel: $a_* = 0.99$ and $P = 0.01$. 
Right panel: $a_* = 1.01$ and $P = -0.01$. 
See text for details.}
\label{f1}
\vspace{1cm}
\includegraphics[type=pdf,ext=.pdf,read=.pdf,width=6cm]{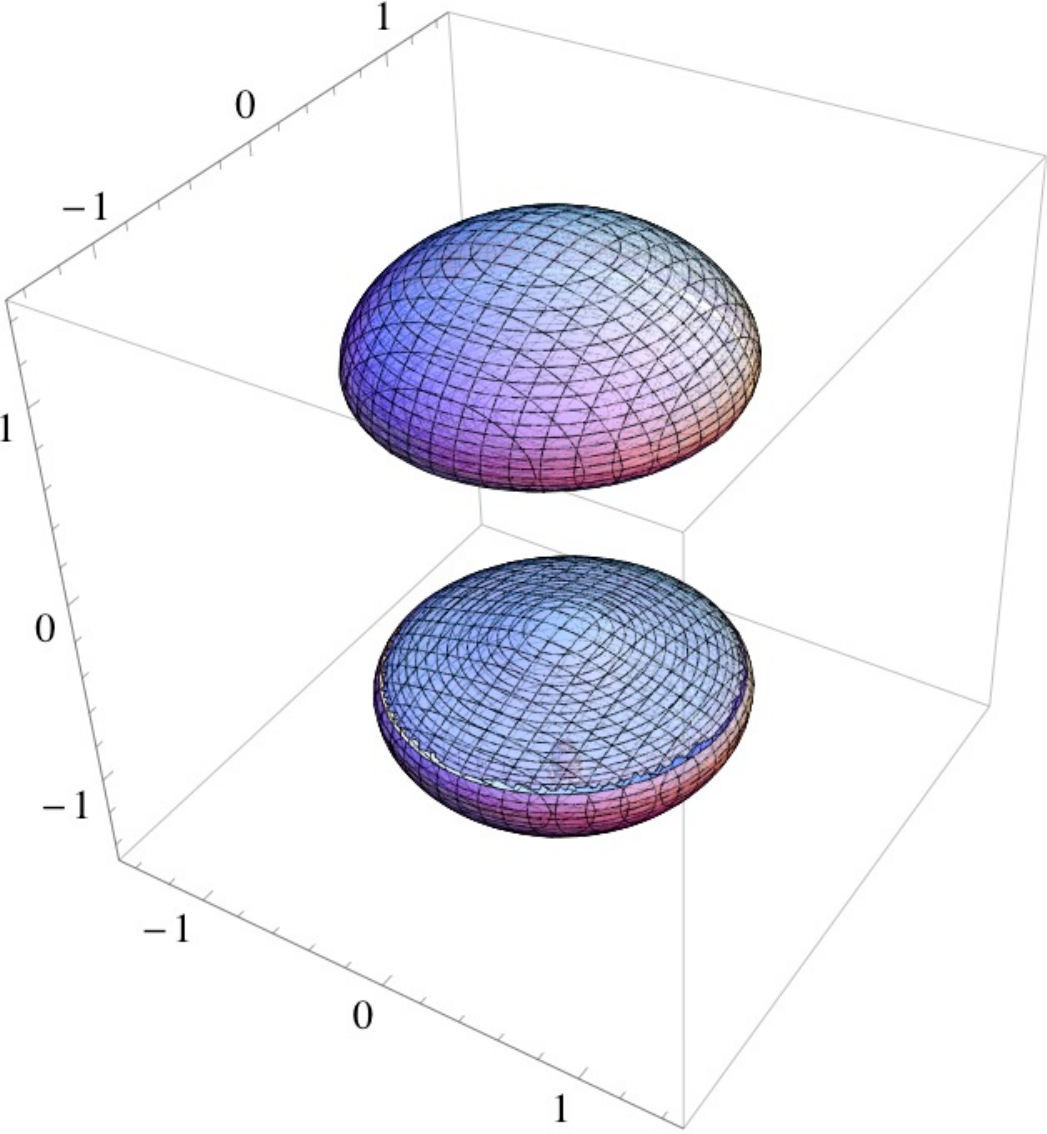}
\hspace{1cm}
\includegraphics[type=pdf,ext=.pdf,read=.pdf,width=6cm]{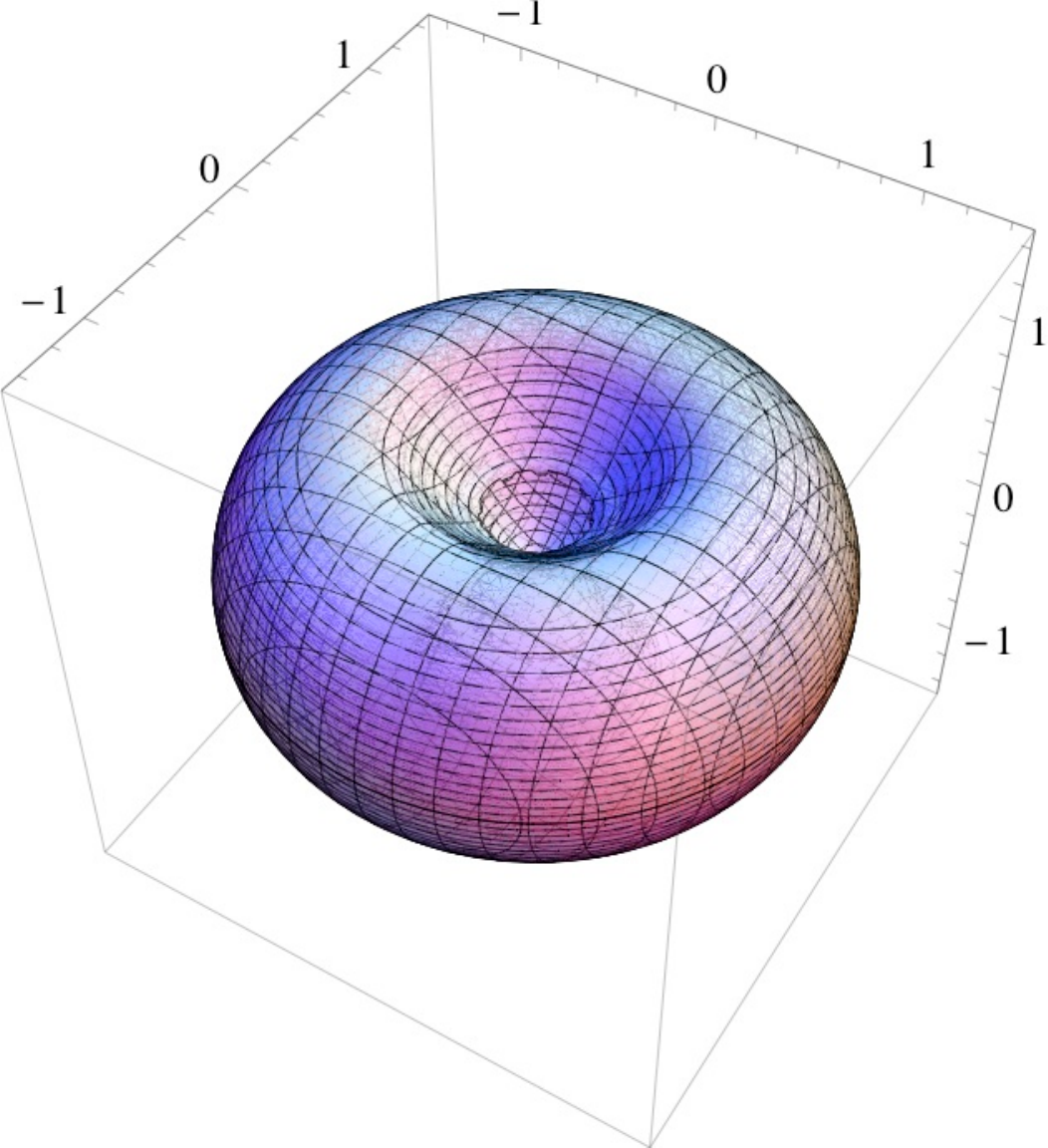}
\par
\vspace{0.3cm} 
\caption{Event horizons of black holes in possible alternative
theories of gravity. 
Left panel: $a_* = 0.9$ and $\epsilon_3 = 1$. 
Right panel: $a_* = 1.1$ and $\epsilon_3 = - 1$. 
See text for details.}
\label{f2}
\end{center}
\end{figure*}

The event horizon is defined by the condition\footnote{Let us notice 
that there are a few definitions of horizon. 
Here we are interested in the {\it event horizon}, i.e. a boundary 
in the space-time beyond which events cannot affect an outside 
observer. In a stationary space-time, the event horizon is also an 
{\it apparent horizon}, which is a surface of zero expansion for a 
congruence of outgoing null geodesics orthogonal to the surface. 
This means that at the apparent horizon null geodesics must have
$dr/dt = 0$, which implies $g^{rr} = 0$, see e.g. Ref.~\cite{poisson}. 
The horizon relevant for the black hole 
thermodynamics is instead the {\it Killing horizon}, which is a null 
hyper-surface on which there is a null Killing vector field. For the 
metric in~(\ref{eq-metric}), the Killing horizon is defined by $g_{tt} 
g_{\phi\phi} - g_{t\phi}^2 = 0$. When the Hawking's rigidity theorem 
can be applied (like in the Kerr space-time), the event horizon 
and the Killing horizon coincide~\cite{hawking}. However, in 
general that is not true.}
\be\label{eq-h1}
\rho^4 \Delta + a^2 \sin^2\theta(\rho^2 + 2 M P r)^2 = 0 \, .
\ee
When $P = 0$, one recovers the classical Kerr result
\be
r_H = M \pm \sqrt{M^2 - a^2} \, ,
\ee
where the sign $+$ and $-$ are respectively for the outer and 
inner horizon. The key-point of the Kerr metric is that
the radial coordinate of the two horizons does not depend on 
$\theta$ and there are two topologically spherical 
horizons for $a < M$, one horizon in the extreme case $a = M$, 
and there is no horizon for $a > M$ (naked singularity). For 
$P \neq 0$, Eq.~(\ref{eq-h1}) depends on $\theta$ and between the cases
of two topologically spherical horizons for low values of $a$
and no horizons for high values of $a$, we find that the two
horizons merge into a horizon with non-trivial spatial 
topology. Let us call $a_*^c$ the lowest value of the 
dimensionless spin parameter for which the BH has a
topologically non-trivial event horizon.
For $P > 0$, there are two disconnected horizons 
with spherical topology, see the left panel in Fig.~\ref{f1}. 
For $P < 0$, there are two disconnected horizons with toroidal 
topology: the one coming from the merger of the outer and inner 
horizons and shown in the right panel of Fig.~\ref{f1}, and 
another small horizon near the origin. The latter is not shown in
the figure and present some peculiar features. For instance, 
while the large toroidal horizon disappears above some critical 
spin parameter, like the horizons of the case $P \ge 0$, the small 
one does not and exists even when $a/M \gg 1$.
Here topologically non-trivial event horizons 
are possible because Eq.~(\ref{eq-metric}) is not a solution of 
the Einstein's equations. Equivalently, the metric 
in Eq.~(\ref{eq-metric}) can be seen as a solution of the 
Einstein's equations in presence of a non-zero effective 
energy-momentum tensor violating the dominant energy 
condition~\cite{bh1}.

\section{Black holes in alternative theories of gravity}

In Ref.~\cite{dim}, the authors have proposed a phenomenological
metric to test gravity in the strong field regime. In other words,
it is not a solution in any known gravity theory, but it is a simple
parametrization of (hopefully generic) deviations from the
Kerr geometry. The metric was obtained by starting from 
a deformed Schwarzschild solution and then by applying
a Newman-Janis transformation. The non-zero metric
coefficients in Boyer-Lindquist coordinates are~\cite{dim}
\be\label{eq-ps}
g_{tt} &=& - \left(1 - \frac{2 M r}{\rho^2}\right) (1 + h)
 \, , \nonumber\\
g_{t\phi} &=& - \frac{2 a M r \sin^2\theta}{\rho^2} 
(1 + h) \, , \nonumber\\
g_{\phi\phi} &=& \sin^2\theta \left[r^2 + a^2
+ \frac{2 a^2 M r \sin^2\theta}{\rho^2} \right] + \nonumber\\
&& + \frac{a^2 (\rho^2 + 2 M r) \sin^4\theta}{\rho^2} 
h \, , \nonumber\\
g_{rr} &=& \frac{\rho^2 (1 + h)}{\Delta + 
a^2 h \sin^2\theta } \, \,\,,\, \,\,\,%\nonumber\\
g_{\theta\theta} = \rho^2 \, ,
\ee
where
\be
\rho^2 &=& r^2 + a^2 \cos^2\theta \, , \nonumber\\
\Delta &=& r^2 - 2 M r + a^2 \, , \nonumber\\
h &=& \sum_{k = 0}^{\infty} \left(\epsilon_{2k} 
+ \frac{M r}{\rho^2} \epsilon_{2k+1} \right)
\left(\frac{M^2}{\rho^2}\right)^k \, .
\ee
The metric has an infinite number of free parameters
$\epsilon_i$ and the Kerr solution is recovered when all
these parameters are set to zero.

The event horizon of a BH described by the metric~(\ref{eq-ps}) 
is given by (please notice that in Ref.~\cite{dim} the authors
use an incorrect definition of event horizon and actually they
compute the Killing horizon)
\be\label{eq-h2}
\Delta + a^2 h \sin^2 \theta = 0 \, .
\ee
Like in~(\ref{eq-h1}), Eq.~(\ref{eq-h2}) depends on $\theta$
and in general, for high values of the spin parameter, the inner 
and the outer horizons merge together, with the result of forming 
a BH with a horizon with non-trivial topology. As a specific
example, we can take the case discussed in~\cite{dim} with
a single free parameter, $\epsilon_3$, and $\epsilon_i = 0$
for $i \neq 3$. However, the picture is qualitatively the same if
we take, for instance, $\epsilon_4$ or $\epsilon_5$ as free
parameter instead of $\epsilon_3$.
For $\epsilon_3 > 0$, the BH turns out to be more
prolate than the Kerr one and, for high spin parameters, one
finds two disconnected horizons with spherical topology, see
the left panel in Fig.~\ref{f2}. If $a_* > 1$, 
there is no horizon, like in the Kerr metric.
For $\epsilon_3 < 0$, the object is more oblate than a Kerr BH 
and one finds that fast-rotating objects have a horizon that looks 
toroidal. It is not really a torus because at $r=0$ there is a naked
singularity which is still connected to the horizon. So, unlike a
doughnut, here there is not a central hole. The case $a_* = 1.1$ and 
$\epsilon_3 = -1$ is shown on the right panel of Fig.~\ref{f2}.
Unlike for the case $\epsilon_3 \ge 0$, when $\epsilon_3 <0$
the horizon never disappears, even for $a/M \gg 1$. 
It just becomes more and more thin and looks like a disk.

\section{Astrophysical black holes}

In the previous sections, we have discussed two specific examples
of spinning BHs in four dimensions
with topologically non-trivial event horizons.
However, in both cases it is
required that the value of spin parameter $a$ exceeds some critical
value. It is thus not clear if such a condition can be satisfied
for astrophysical BHs.
For instance, in the Kerr case, the solution describes a BH for 
$a \le M$ and a naked singularity for $a>M$. However, 
Kerr naked singularities can unlikely be of astrophysical interest,
as it is apparently impossible to overspin a Kerr BH up to 
$a > M$~\cite{eb}. So, the purpose of this section is to show
that BHs with topologically non-trivial event horizon could be created in
the Universe.

A natural and very efficient mechanism to spin a compact object
up is through the process of gas accretion from a disk. One can 
assume that the disk is on the equatorial plane of the object 
and that the disk's inner edge is located at a radius $r_{in}$.
When the gas reaches the inner edge, it plunges to the BH
with no further emission of radiation. If at the radius 
$r_{in}$ the specific energy $E = - u_t$ and the specific angular 
momentum $L = u_\phi$ of a gas particle with 4-velocity $u^\mu$ are 
respectively $E_{in}$ and $L_{in}$, the compact object changes 
its mass $M$ and its spin angular momentum $J$ by
\be
\delta M &=& E_{in} \delta m \, , \nonumber\\ 
\delta J &=& L_{in} \delta m \, ,
\ee
where $\delta m$ is the gas rest-mass. The evolution of the spin 
parameter is thus governed by the following equation~\cite{bardeen}
\be\label{eq-a}
\frac{da_*}{d\ln M} = \frac{1}{M} 
\frac{L_{in}}{E_{in}} - 2 a_* \, .
\ee
$E_{in}$ and $L_{in}$ depend on the metric
of the space-time (see Eq.~(\ref{e-l}) below) 
and on the model of the accretion disk.

The simplest case is the geometrically thin and optically thick
disk~\cite{noth}, whose inner edge is at the marginally stable 
circular orbit (also called innermost stable circular orbit, or 
ISCO): $r_{in} = r_{ms}$. For a generic stationary and axisymmetric 
space-time, the calculation of $E_{ms}$ and $L_{ms}$ goes as 
follows (see e.g. Appendix~B in Ref.~\cite{bb} for more details). 
The disk's gas moves on nearly geodesic circular orbits
and the equations of motion are
\be
\dot{t} &=& \frac{E g_{\phi\phi} + 
Lg_{t\phi}}{g_{t\phi}^2 - g_{tt}g_{\phi\phi}} \, , \\
\dot{\phi} &=& - \frac{E g_{t\phi} + 
Lg_{tt}}{g_{t\phi}^2 - g_{tt}g_{\phi\phi}} \, , \\
g_{rr}\dot{r}^2 + g_{\theta\theta}
\dot{\theta}^2 &=& V_{\rm eff}(E,L,r,\theta) \, ,
\ee
where $V_{\rm eff}$ is the effective potential 
\be
V_{\rm eff} = \frac{E^2 g_{\phi\phi} + 2ELg_{t\phi}
+ L^2g_{tt}}{g_{t\phi}^2 - g_{tt}g_{\phi\phi}} - 1 \, .
\ee
Circular orbits in the equatorial plane are located at the
zeros and the turning points of the effective potential:
$\dot{r}=\dot{\theta}=0$ implies $V_{\rm eff} = 0$, and
$\ddot{r}=\ddot{\theta}=0$ requires $\partial_rV_{\rm eff}
= \partial_\theta V_{\rm eff} =0$. $E$ and $L$ turn out to be
\be\label{e-l}
E &=& - \frac{g_{tt} + g_{t\phi}\Omega}{\sqrt{-g_{tt} 
- 2 g_{t\phi}\Omega - g_{\phi\phi}\Omega^2}} \, , \nonumber\\
L &=& \frac{g_{t\phi} + g_{\phi\phi}\Omega}{\sqrt{-g_{tt} 
- 2 g_{t\phi}\Omega - g_{\phi\phi}\Omega^2}} \, ,
\ee
where 
\be
\Omega = \frac{d\phi}{dt} = \frac{-\partial_r g_{t\phi} \pm 
\sqrt{(\partial_r g_{t\phi})^2 - (\partial_r g_{tt})
(\partial_r g_{\phi\phi})}}{\partial_r g_{\phi\phi}}
\ee
is the orbital angular velocity and the sign $+$
($-$) is for corotating (counterrotating) orbits. The orbits 
are stable under small perturbations if $\partial_r^2V_{\rm eff} 
\le 0$ and $\partial_\theta^2V_{\rm eff} \le 0$. At the ISCO,
either $\partial_r^2V_{\rm eff} = 0$ or $\partial_\theta^2
V_{\rm eff} = 0$. In this way, one determines $E_{in} = E_{ms}$ 
and $L_{in} = L_{ms}$ and can integrate Eq.~(\ref{eq-a}) to get 
the equilibrium spin parameter $a_*^{eq}$\footnote{In the case 
of non-Kerr background, the picture may be more complicated
and, in some cases, the gas may not be able to plunge from the 
ISCO to the BH, see Ref.~\cite{bb2}. If this is the case,
accretion is possible only if the gas loses additional energy 
and angular momentum. However, such an effect
occurs only for ``extreme'' objects, typically with
spin parameter $a_*$ well above the equilibrium value 
$a_*^{eq}$.}.
For instance, an initially non-rotating BH in 
General Relativity reaches the equilibrium spin parameter 
$a_*^{eq} = 1$ after having increased its mass by a 
factor $\sqrt{6} \approx 2.4$~\cite{bardeen}. If the compact 
object is not a Kerr BH, the final value of the spin 
parameter may be larger than 1
and super-spinning compact objects may be created~\cite{esp,super}.

In the case of non-Kerr BHs, topologically 
non-trivial horizons may have astrophysical relevance if the
deviations from the Kerr geometry are 
not too small. For a loop BH and $P = 0.01$, the 
equilibrium spin parameter is $a_*^{eq} \approx 0.9831$, while 
the transition of the topology of the horizon happens at 
$a_*^{c} \approx 0.9805$. A few other cases are reported
in Tab.~\ref{t-eq} and we have checked that all the results
do not depend on the ambiguities related to $\Sigma$.
As $|P| \rar 0$, $a_*^{eq}$ and $a_*^{c}$ go to 1 and even if 
$a_*^{eq}$ is always larger than $a_*^{c}$, the difference is 
smaller and smaller. This fact forbids the possibility of
changing the topology of the horizon for small values of $|P|$,
 as $a_*^{eq}$
cannot be reached in the reality. In particular, 
the radiation emitted by the disk and
captured by the BH reduces the value 
of $a_*^{eq}$ computed from Eq.~(\ref{eq-a}), as the radiation
with angular momentum opposite to the BH spin has
larger capture cross section. For example, in the case of the
Kerr metric one finds the well-known ``Thorne's limit''
$a_*^{eq} \approx 0.998$~\cite{th}. To get a rough estimate of the 
effect, we can assume that the radiation captured by the BH
reduces $a_*^{eq}$ by 0.002 (actually for loop BHs the effect
is smaller, because for $P \neq 0$ the ISCO 
radius at $a_*^{eq}$ is larger than the Kerr one). In this
case, the accretion process from a thin disk can spin the BH
up to the critical value $a_*^c$ and induces the topology
transition only if $P \gtrsim 0.01$ or $P \lesssim - 0.001$.

Tab.~\ref{t-eq2} shows the same quantities for the BHs described
by the metric~(\ref{eq-ps}) (see also Fig.~2 in Ref.~\cite{dim},
where the authors show $a_*^c$ as a function of $\epsilon_3$
for $\epsilon_3 > 0$). 
Even here, $a_*^{eq} > a_*^c$,
but the difference is smaller and smaller as $\epsilon_3 \rar 0$.
Like for the loop BHs, the topology transition of the event
horizon occurs at $a_* < 1$ for more prolate objects, and at 
$a_* = 1$ for more oblate objects.

\begin{table}
\begin{center}
\begin{tabular}{c c c c c}
\hline \\
$P$ & \hspace{.5cm} & $a_*^{eq}$ & 
\hspace{.5cm} & $a_*^c$ \\ \\
\hline 
$0.01$ & & 0.9831 & & 0.9805 \\
$0.001$ & & 0.9981 & & 0.9980 \\ 
$0.0001$ & & 0.9998 & & 0.9998 \\
$0.0$ & & 1.0 & & --- \\
$-0.0001$ & & 1.0002 & & 1.0000 \\
$-0.001$ & & 1.0020 & & 1.0000 \\ 
$-0.01$ & & 1.0192 & & 1.0000 \\
\hline
\end{tabular}
\end{center}
\caption{Black holes inspired by Loop Quantum Gravity.
Equilibrium spin parameter, $a_*^{eq}$, and critical 
spin parameter separating black holes with topologically 
different horizons, $a_*^c$, for some values of the polymeric 
function $P$. The case $P=0$ corresponds to the classical Kerr
metric.}
\label{t-eq}
\end{table}

\begin{table}
\begin{center}
\begin{tabular}{c c c c c}
\hline \\
$\epsilon_3$ & \hspace{.5cm} & $a_*^{eq}$ & 
\hspace{.5cm} & $a_*^c$ \\ \\
\hline 
$10.0$ & & 0.5608 & & 0.4355 \\
$1.0$ & & 0.8705 & & 0.7910 \\ 
$0.1$ & & 0.9735 & & 0.9596 \\
$0.0$ & & 1.0 & & --- \\
$-0.1$ & & 1.0334 & & 1.0000 \\
$-1.0$ & & 1.1854 & & 1.0000 \\ 
$-10.0$ & & 1.6531 & & 1.0000 \\
\hline
\end{tabular}
\end{center}
\caption{Black holes in possible alternative theories of gravity.
Equilibrium spin parameter, $a_*^{eq}$, and critical 
spin parameter separating black holes with topologically 
different horizons, $a_*^c$, for some values of the parameter 
$\epsilon_3$. The case $\epsilon_3=0$ corresponds to the Kerr
metric.}
\label{t-eq2}
\end{table}

\section{Observational constraints}

Up to now, we have considered arbitrary deviations from the Kerr
geometry. However, either $P$ for the metric~(\ref{eq-metric})
and $\epsilon_3$ for~(\ref{eq-ps}) are subject to constraints
from current observations.

For the loop BHs, one can apply the Birkhoff's theorem and
constrain $P$ by using current observational data
in the Solar System. In the parametrized post-Newtonian (or PPN)
framework~\cite{will}, the line element of the asymptotic 
space-time in spherical coordinates is
\be
ds^2 = - A(r) dt^2 + B(r) dr^2 + r^2d\Omega^2 \, ,
\ee
where
\be
A(r) &=& 1 - \frac{M}{r} 
+ 2 (\beta_{PPN} - \gamma_{PPN})\frac{M^2}{r^2} + . . . \, , \\
B(r) &=& 1 + 2 \gamma_{PPN} \frac{M}{r} + . . . \, ,
\ee
and $\beta_{PPN}$ and $\gamma_{PPN}$ are the PPN parameters. In classical
General Relativity $\beta_{PPN} = \gamma_{PPN} = 1$ and the strongest constraints
come from the Lunar Laser Ranging experiment~\cite{llr} and 
the Cassini spacecraft~\cite{cassini}
\be
|\beta_{PPN} - 1| &<& 2.3 \cdot 10^{-4} 
\quad ({\rm LLR}) \, , \nonumber\\
|\gamma_{PPN} - 1| &<& 2.3 \cdot 10^{-5} 
\quad ({\rm Cassini}) \, .
\ee
The asymptotic form of the loop-modified Kerr metric reads
\be
-g_{tt} &=& 1 - \frac{2m}{r} + 2 
\frac{4 (P - P^2 + P^3)}{(1-P)^4} \frac{m^2}{r^2} + . . . \, , 
\nonumber\\
g_{rr} &=& 1 + 2 \frac{(1+P)^2}{(1-P)^2} \frac{m}{r} + . . . \, ,
\ee
where $m = (1 - P)^2 M$ is the gravitational mass as measured
by a distant observer. From the Cassini spacecraft, we get
\be\label{ccc}
|P| < 0.6 \cdot 10^{-5} \, .
\ee
So, if $P$ is really a constant, it must be so small that the 
accretion process can unlikely create loop BHs with non-trivial 
topology. However, the Immirzi parameter may be a running 
constant, approaching 0 at low energies/large distances and 
1 at high energies/short distances, as discussed in~\cite{run}. 
If this is the case, the accretion process might still create BHs 
with non-trivial topology, as the constraint~(\ref{ccc}) would 
hold only far from the BH.

Let us now consider the metric~(\ref{eq-ps}).
As discussed in Ref.~\cite{dim}, the Newtonian limit requires
$\epsilon_0 = \epsilon_1 = 0$ and the Lunar Laser Ranging
experiment demands $|\epsilon_2| < 4.6 \cdot 10^{-4}$.
On the other hand, there are no bounds on $\epsilon_i$ for
$i \ge 3$ from the Solar System. Following the argument 
in Ref.~\cite{agn}, it is possible to constrain the deformation
parameters with $i \ge 3$ 
from the estimate of the mean radiative efficiency of
active galactic nuclei (AGN). Let us notice, however, that the
final bound has to be taken with some caution.
One can notice that the most luminous 
super-massive objects in galactic nuclei have a radiative efficiency 
$\eta > 0.15$~\cite{elvis}. 
There are several sources of uncertainty to
get this bound, but this value seems to be a reliable lower
limit. Assuming that $\epsilon_i$ do not depend on 
$M$ or $J$, we can constrain possible deviations from
the Kerr metric, as $\eta \le 1 - E_{ms}$. In the 
specific case $\epsilon_3 \neq 0$ and $\epsilon_i = 0$ for 
$i \neq 3$, one finds:
\be
-1.1 < \epsilon_3 < 25 \, .
\ee
Such a bound is weak and surely does not forbid the possibility
of astrophysical BHs with non-trivial topology.

\section{Conclusions}

In 4-dimensional General Relativity, a stationary BH must 
have a spherical horizon, while a toroidal horizon is allowed for a
very short time. It is thus thought that the spatial topology of the 
horizon of astrophysical BHs is a 2-sphere. However, if the
current BHs candidates are not the BHs predicted by General 
Relativity, this conclusion may be wrong. Here we have discussed
two examples of 4-dimensional non-Kerr spinning BHs and we have shown 
that for high values of the spin parameter the topology of the
event horizon of these objects can change. Unfortunately,
our current knowledge of these objects
is definitively limited. Here we have considered the loop-improved
Kerr metric found in Ref.~\cite{bh2} and the phenomenological
metric proposed in~\cite{dim} to perform tests of strong gravity.
Interestingly,
they present quite remarkable and qualitatively similar features 
in the case of fast-rotating objects and we guess that these
properties may be common for non-Kerr spinning BHs.
In particular, it seems that BHs more prolate than the Kerr one
develop two disconnected topologically spherical horizons 
above some critical spin parameter. In the case of fast-rotating
objects more oblate than a Kerr BH, their horizon looks more
like a torus, even if the central hole may be closed (like in the
case of the BH in the right panel of Fig.~\ref{f2}, in which the event 
horizon extends up to the central singularity at $r=0$). In both
this examples, the accretion process may overspin these
objects above the critical spin parameter and induce the
topology transition of the horizon. The topology change does not
happen as a result of some jump or tunneling, but this fact
should not be seen suspiciously: even in numerical 
simulations in General Relativity the gravitational collapse
can produce a toroidal BH and then the hole quickly closes up
continuously, as found for the first time in~\cite{teuk}.

As final remark, let us notice that here 
we have not discussed the stability of
these BHs. However, this issue cannot be addressed for the
metrics~(\ref{eq-metric}) and (\ref{eq-ps}): we do not know the
field equations of the gravity theory having Eqs.~(\ref{eq-metric}) 
and (\ref{eq-ps}) as solution, and therefore we cannot predict 
the evolution of small perturbations on these backgrounds.
On very general grounds, we can simply say the event horizon
of very fast-rotating objects becomes likely too small to prevent
the ergoregion instability~\cite{ergo}. However, such an instability
may occur only for BHs with spin parameters $a_* > a_*^{eq}$, 
which would be anyway unstable configurations. If, on the
contrary, the instability appears at $a_* < a_*^{eq}$, as the 
spin-up due to the accretion process is an unavoidable phenomenon,
these BHs would be a source of gravitational waves, potentially 
detectable by future experiments.

%%%%%%%%%%%%%%%%%%%%%%%%%%%%%%%

\begin{acknowledgments}
We would like to thank Alexander Dolgov and Piero Nicolini
for critically reading a preliminary version of this manuscript and
providing useful feedback. We are also grateful to Francesco
Caravelli for helpful comments. The work of C.B. was supported by 
World Premier International Research Center Initiative (WPI Initiative), 
MEXT, Japan, and by the JSPS Grant-in-Aid for Young Scientists 
(B) No. 22740147. Research at Perimeter Institute is supported by 
the Government of Canada through Industry Canada and by the Province 
of Ontario through the Ministry of Research \& Innovation. L.M. 
would like to acknowledge support and hospitality from the Institute 
for the Physics and Mathematics of the Universe at The University of 
Tokyo, where this work was started. C.B. would like to acknowledge 
support and hospitality from the Perimeter Institute, where this work 
was finalized.
\end{acknowledgments}

%%%%%%%%%%%%%%%%%%%%%%%%%%%%%%

\end{document}